\documentclass[12pt]{JHEP3}

\newcommand{\be}{\begin{equation}}
\newcommand{\ee}{\end{equation}}
\newcommand{\bea}{\begin{eqnarray}}
\newcommand{\eea}{\end{eqnarray}}
\newcommand{\ba}{\begin{array}}

\newcommand{\ea}{\end{array}}

\def\bbox{{\,
\lower0.9pt\vbox{\hrule \hbox{\vrule height 0.2 cm
\hskip 0.2 cm \vrule height 0.2 cm}\hrule}\,}}
\newcommand{\dsl}{\pa \kern-0.5em /}

\newcommand{\nn}{\nonumber \\}

\def\xxx{{\tt hep-th}/}

\def\Tr{{\rm Tr\,}}

\def\ds{\raise.15ex\hbox{/}\kern-.57em\partial}
\def\Ds{\,\raise.15ex\hbox{/}\mkern-13.5mu D}
\def\Tr{{\rm Tr}\,}
\def\n{\nonumber}
\newcommand{\beq}{\begin{equation}}
\newcommand{\eeq}{\end{equation}} \newcommand{\beqn}{\begin{eqnarray}}
\newcommand{\eeqn}{\end{eqnarray}}

\preprint{
KIAS-P02081\\
ITFA-2003-03\\
hep-th/0301120\\
}

\title{\Large\bf Instantons of M(atrix) Theory in PP-Wave Background}

\author{Jung-Tay Yee$^{1,2}$ and Piljin Yi$^{1}$\\
jungtay@science.uva.nl and piljin@kias.re.kr\\
\vskip 0.2mm
$^{1}$School of Physics, Korea Institute for Advanced Study\\
207-43, Cheongryangri-Dong, Dongdaemun-Gu, Seoul 130-012, Korea\\
\vskip 0.2mm
$^{2}$Institute for Theoretical Physics, University of Amsterdam \\
Valckenierstraat 65, 1018 XE Amsterdam, The Netherlands}

\abstract{M(atrix) theory in PP-wave background possesses a
discrete set of classical  vacua, all of which
preserves 16 supersymmetry and interpretable as collection of
giant gravitons.  We find Euclidean instanton solutions that
interpolate between them, and analyze their properties. 
Supersymmetry prevents direct mixing between different vacua 
but still allows effect of instanton to show up in higher order 
effective interactions, such as analog of $v^4$ interaction of 
flat space M(atrix) theory. An explicit construction of zero modes
is performed, and Goldstone zero modes, bosonic and fermionic,
are identified. We further generalize this to  massive M(atrix) 
theory that includes fundamental hypermultiplets, corresponding 
to insertion of longitudinal fivebranes in the background. After 
a brief comparison to their counterpart in $AdS\times S$, we 
close with a summary.}

\begin{document}

\section{Introduction}

Maximally supersymmetric plane waves background in eleven
dimensions is obtained from either $AdS_7 \times S^4$ or $AdS_4
\times S^7$ by taking the Penrose limit \cite{ppwaves}. String theory
or M-theory in the plane waves background \cite{bmn} admits rich
tractable structures which are useful to understand AdS/CFT
duality and physics in the presence of Ramond-Ramond flux:
Lightcone Green-Schwarz action of type IIB string theory is free
and exactly quantizable \cite{metsaev}. M-theory in eleven
dimension has a M(atrix) theory formulation with a
mass parameter $\mu$ which can be used as an expansion
parameter in appropriate combination with usual Yang-Mills
coupling \cite{DSJvR,plefka}.

For M(atrix) theory in the plane waves background \cite{bmn},
first written down by Berenstein, Maldacena, and Nastase (BMN),
the vacua are characterized by discrete set of fuzzy spheres
which preserve all 16 dynamical supersymmetries. They are
nothing but D0-branes expanded by Myers' dielectric term
\cite{myers}. Tracing back to $AdS \times S$ space, 
these fuzzy spheres originate from giant gravitons rotating
with the velocity of light in the $AdS$ space or in the sphere 
\cite{gg,kimyee,gginpp}.  
One purpose of this note is to examine physics of giant gravitons 
in this  more tractable quantum theory.

Given such  degenerate set of vacua, an obvious thing
to ask is whether there are instantons that interpolate between 
different classical vacua. It turns out that search for instantons
is greatly simplified thanks to BPS-like properties of the  
latter. That is, instantons interpolating between fully supersymmetric 
vacua of this theory preserves half of ${\cal N}=16$ dynamical
supersymmetry.  Those instantons that interpolate between any 
nontrivial vacuum  and the perturbative
vacuum are particularly simple, it turns out, and an explicit 
counting of zero modes, bosonic and fermionic, is carried out.
As supersymmetry may already suggests, these instantons have too
many fermionic Goldstone zero modes to mix vacua, and will 
contribute to higher order interaction terms only.

This note is organized as follows. After a brief review of  
the BMN M(atrix) theory in section 2, we take analytic continuation  
into the Euclidean time in section 3.  Here we have to deal with subtlety 
associated with Majorana condition on fermions, utilizing ``double 
fermion technique'' \cite{euclideanaction}.  BPS-like equation 
follows then, either by writing Euclidean action in a complete square form
or by writing down supersymmetry condition. We solve for all instantons
that interpolate between the perturbative vacuum and arbitrary 
collections of giant gravitons.

In section 4, we construct bosonic and fermionic zero modes around
these instantons explicitly. Apart from 8 fermionic zero modes
and a single bosonic zero mode, each arising from spontaneously
broken symmetry, we find a large number of extra zero modes for
most of these instantons. This leads us to believe that there are
more diverse form of instantons, which we confirm in part in section 5.  

Section 6 repeats the computation for certain mass-deformed 
${\cal N}=8$ SYQM \cite{kimleeyi}. 
The latter have been discovered and identified  as M(atrix) 
theory in the presence of longitudinal fivebranes . 
Among
the vacua of this latter theory are giant gravitons trapped by
the longitudinal fivebranes. We again isolate and study supersymmetric
instantons that  interpolate from a generic vacuum to the perturbative 
one. After a review of giant gravitons and instantons in $AdS\times S$, 
the Penrose limit thereof in section 7, we close with summary.

\section{BMN Matrix Theory and Classical Vacua}

We start with a summary of the mass deformed BMN M(atrix) theory 
\cite{bmn}. Here we review some basic
properties of the theory, in part to set up the notation, and
recall its classical vacuum structure.

\subsection{Lagrangian and Supersymmetry}

Introducing $16\times 16$ real and symmetric
$SO(9)$ Dirac matrices,
\beq
\gamma_I \qquad (I=1,2...9)
\eeq
the Lagrangian of the mass-deformed M(atrix) theory of BMN is
\beqn
L&=&  \frac12\Tr \left( \sum_I (D_0X_I)^2 +
\frac{1}{2}\sum_{IJ}[X_I,X_J]^2 + i\Psi^T D_0\Psi
-\Psi^T \gamma_I[X_I, \Psi ] \right)\nn
&+&\frac{1}{2}\Tr\left( -\left(\frac{\mu}{3}\right)^2
\sum_{a=5,6,7} (X_a)^2 -\left(\frac{\mu}{6}\right)^2
\sum_{s=1,2,3,4,8.9}(X_s)^2 \right) \n \\
&+&\frac{1}{2}\Tr\left( \frac{2i\mu}{3}
\epsilon_{abc} X_a X_b X_c -\frac{i\mu}{4} \Psi^T\gamma_{567}\Psi
  \right),
\eeqn
Here the fermion $\Psi$ is Majorana.\footnote{The
Majorana condition in 9+1 dimensions is achieved by imposing a
Majorana condition,
\beq
\Psi=C\bar \Psi^T
\eeq
for some charge conjugation matrix $C$. For 16-component fermion here, 
we may introduce $B$ which is a $16\times 16$ block off $C(\Gamma^0)^T$ 
such that
\beqn
\Psi=B\Psi^*,\qquad
B(\gamma^I)^*B^{-1}=\gamma^I, \qquad
BB^*=1
\eeqn
In the present case, we  may choose $B=1$ so that $\gamma^I$'s are real and
symmetric.}
In addition, both $X^I$ and $\Psi$ are valued in  $N\times N$
hermitian matrices for an integer $N\ge 1$. Dimensionful
quantity $\mu$ introduces a mass-gap to the system and make the
spectrum of the theory completely discrete. For the most part, we will
be concerned with ground state sector of this theory.

This Lagrangian was shown to be invariant under the following
16 (dynamical) supersymmetry transformation
\beqn
\delta A_0 &=& i\Psi^T \epsilon ,   \n \\
\delta X_I&=& i\Psi^T \gamma_I \epsilon , \n \\
\delta \Psi &=& \left( D_0 X_I \gamma_I  -\frac{i}{2}
[X_I,X_J]\gamma_{IJ} \right)\epsilon \n\\
&+&\left(\frac{\mu}{3}\sum_{a=5,6,7} X_a\gamma_a
\gamma_{567} -\frac{\mu}{6}\sum_{s=1,2,3,4,8,9} X_s\gamma_s
\gamma_{567}
\right) \epsilon.
\eeqn
provided that we force the following explicit time-dependence of the
transformation parameter $\epsilon$,
\beq
\epsilon(t) = e^{-\frac{\mu}{12}\gamma_{567}t} \epsilon_0 , \label{time}
\eeq
which makes the superalgebra quite unconventional. In particular,
supercharges do not commute with the Hamiltonian, but raise
or lower energy by $\mu/12$ unit. The notion of ``supermultiplet''
no longer implies degeneracy in this deformed superalgebra.

\subsection{Classical Vacua are Collections of Giant Gravitons}

The potential of the mass-deformed matrix theory of BMN lifts 
flat directions altogether, and instead leave behind a set of
discrete vacua. One way to see this is to realize that  
the potential can be written in a complete square form. 
Setting $X^i=0$ for $i=1,2,3,4,8,9$ as a matter of convenience, 
the potential reduces to 
\beqn
V\rightarrow \Tr \left(-\frac{1}{4}\sum_{a,b=5,6,7}[X^a, X^b]^2 +
\frac12 \left(\frac{\mu}{3}\right)^2 \sum_{a=5,6,7} (X^a)^2
-i\frac{\mu}{3} \epsilon_{abc} X^a X^b X^c \right), \label{V}
\eeqn 
which can be made into a complete square, 
\beqn 
-\frac14 \sum
\Tr\left( [X^b,X^c] +\frac{i\mu}{3} \epsilon_{bca}X^a\right)^2.
\eeqn 
Thus zero of the potential is given by $X^a$'s satisfying a
set of commutation relationship 
\beq
[X^a,X^b]=-\frac{i\mu}{3}\epsilon^{abc}X^c ,
\eeq 
and classical vacua
are given by static configurations of the form 
\beqn
X^a=-\frac{\mu}{3}J^a , 
\eeqn 
with any $N$ dimensional
representation $J^a$ of $SU(2)$.

Nothing here requires $J$'s to form an irreducible representation,
so we should look for general $N$ dimensional representation. We
may do this by filling out three $N\times N$ matrices with various
irreducible representations of $SU(2)$. Different ordering
among such irreducible blocks are all gauge equivalent, and
therefore classification of vacua comes down to classifying
inequivalent partitions of the integer $N$.

These classical vacua are reminiscent of bound states of 
original M(atrix) theory \cite{Yi}. In the latter, the bound 
states are supposed to exist for any number of D0's, which
means that Hilbert space contain sectors of $n_p$ particles bound
together for any partition of $N$ ($ N=\sum_p n_p$). 
Given the $U(N)$ BMN M(atrix) theory, there is precisely one such 
distinct state for each partition $N=\sum_p  n_p$, corresponding 
to a direct sum of $n_p$ dimensional irreducible representations of
$SU(2)$. These classical states (and their quantum counterparts) are
closest thing we have to usual Kaluza-Klein modes of supergraviton, 
or equivalently BPS bound states of D-particles. In fact, one may
think each of these $n_p\times n_p$ blocks as the bound state
$n_p$ D-particles which are blown-up into a spherical membrane due
to Myers' dielectric effect~\cite{myers}. In the 
mass deformed version here, flat directions are lost
and everything is confined near the origin, yet it is tantalizing
that hint of these D0 bound states still survives in the form of
discrete set of classical vacua \cite{gg}.

\section{Euclidean  Instantons}

\subsection{Euclidean Action and Supersymmetry}

We wish to look for interpolating solution between different vacua,
so a natural thing to do is to Euclideanize the theory. The upshot 
is that we can do this effectively by mapping
\beq 
t\rightarrow -i\tau 
\eeq 
which is actually is a result of more careful operation that maintains 
10-dimensional origin for fermions. Rest of this subsection will 
address how to do this more carefully.

Analytic continuation to the Euclidean signature
is often hazardous when real fermions are present. The problem 
with real fermions in field theory is that representation theory 
under $SO(d,1)$ Clifford algebra is different from that under 
$SO(d+1)$.
Since fermions of this quantum mechanics came from $SO(9,1)$
spinors, this sort of problem again shows up when we try to
perform a Euclidean continuation of this quantum mechanics. 
On the other hand, this also tells us that the
trick we can use is already available in field theory language,
where we abandon hermiticity of Lagrangian, 
``double'' fermion content by regarding the
conjugated fermions as independent, and then impose
certain ``Majorana-like'' constraints on the pair 
\cite{euclideanaction}.

Let us start by rewriting the Minkowski action in $SO(9,1)$ spinor
notation.
 \beqn
L_M&=& \frac12\Tr \left( \sum_I (D_0X_I)^2 +
\frac{1}{2}\sum_{IJ}[X_I,X_J]^2 - i\bar\Psi \Gamma^0 D_0\Psi
-\bar\Psi \Gamma_I[X_I, \Psi ] \right)\nn &+&\frac{1}{2}\Tr\left(
-\left(\frac{\mu}{3}\right)^2 \sum_{a=5,6,7} (X_a)^2
-\left(\frac{\mu}{6}\right)^2
\sum_{s=1,2,3,4,8.9}(X_s)^2 \right) \n \\
&+&\frac{1}{2}\Tr\left( \frac{2i\mu}{3}
\epsilon_{abc} X_a X_b X_c -\frac{i\mu}{4} \bar\Psi \Gamma_{567}\Psi
  \right),
\eeqn
where $\Psi$ is Majorana-Weyl spinor and 10 dimensional $\Gamma$ matrices
are given
\beqn
&& \Gamma^0=1 \otimes i\sigma_2 ,\n\\
&& \Gamma_I = \gamma_I\otimes \sigma_1 \;\; (I=1,2...9),\n\\
&& \Gamma^{11} = 1\otimes \sigma_3, \eeqn
where $\gamma_I$ are usual $16\times 16$ hermitian $SO(9)$ Dirac matrices.

Weyl spinor $\Psi$ is subject to the chirality condition
\beqn
 \Gamma_{11} \Psi = - \Psi.
\eeqn which effectively reduces $\Psi$ to be 16-component, 
justifying the fact that the same notation $\Psi$ is used for 
16-component spinor of previous section and for the current 
32-component spinor with upper half identically zero. 
In particular this means that $32\times32$
matrices $\Gamma$'s are effectively $16\times 16$ matrices of the
following form, 
\beqn &&\Gamma^0\rightarrow +1 , \nn
&&\Gamma_I\rightarrow \gamma_I ,
\eeqn 
when acting on $\Psi$
directly. 
Finally, the Majorana condition on $\Psi$ is 
\beqn
 \Psi = C \bar\Psi^T,
\eeqn
where the charge conjugation matrix C is chosen to be
\beqn
 C = 1 \otimes i \sigma_2,
\eeqn
which satisfies
\beqn
 C \Gamma_I = - \Gamma^T C_I, \qquad C^T = -C.
\eeqn
This gives the Minkowski action of the previous section.

We may Euclideanize the action and the supersymmetry by taking
\beqn
 t = -i \tau \quad(\Rightarrow D_0 = i D_\tau),
 \qquad \Gamma^0 = -i \Gamma_{10},
\eeqn
where
\beqn
 \Gamma_{10} = 1 \otimes \sigma_2.
\eeqn
Imposing Euclidean Weyl condition on the fermions, we
have effective reduction of Dirac matrices to $16\times 16$.
\beqn
&&\Gamma_{10}\rightarrow -i  \n \\
&&\Gamma_I\rightarrow \gamma_I \n \\
&&\Gamma_{10} \Gamma_I \rightarrow i \gamma_I,
\eeqn
when acting on $\Psi$ directly.
Furthermore, we use ``fermions doubling'' trick, introducing $\bar \chi$
instead of $\bar\Psi$.
\beqn L_E&=&  \frac12\Tr \left( \sum_I (D_\tau X_I)^2-
\frac{1}{2}\sum_{IJ}[X_I,X_J]^2 + \bar\chi D_{\tau}
\Psi +\bar\chi \gamma_I[X_I, \Psi ] \right)\nn
&+&\frac{1}{2}\Tr\left( +\left(\frac{\mu}{3}\right)^2
\sum_{a=5,6,7} (X_a)^2 +\left(\frac{\mu}{6}\right)^2
\sum_{s=1,2,3,4,8.9}(X_s)^2 \right) \n \\
&+&\frac{1}{2}\Tr\left(- \frac{2i\mu}{3}
\epsilon_{abc} X_a X_b X_c +\frac{i\mu}{4} \bar\chi \gamma_{567}\Psi
  \right).
\eeqn
Without Majorana condition, the path integral measure would be
\beq
[d\Psi][d\bar\chi].
\eeq
However, we must impose a ``Majorana-like'' condition on the pair
\beqn
\bar{\chi} = \Psi^T C,
\eeqn
where $C$ is the same charge conjugation matrix in Minkowski
signature. The measure is effectively,
\beq
[d\Psi].
\eeq
This is analogous to contour integral over complex bosonic 
coordinates.\footnote{
Note also that we have abandoned hermiticity of the Lagrangian in doing 
this. This is not something new in fact. The same happens even for bosons 
when we start with canonical variables. For fermions, in a sense, we
are always working with canonical variables.}

Supersymmetry transformations remain intact once Wick rotation is
performed. Transformation of bosons
are
 \beqn
\delta A_{10} &=&  {1 \over 2} (\bar\chi \epsilon - \bar\eta \Psi) ,   \n \\
\delta X_I&=&  {i \over 2} (\bar \chi \gamma_I \epsilon - \bar \eta \gamma_I
  \Psi) ,
\eeqn
where $\bar\eta$ is related to $\epsilon$ via
\beqn
 \bar \eta = \epsilon^T C \rightarrow - \epsilon^T.
\eeqn
Transformation of fermions are such that,
\beqn
\delta \Psi &=& \left(i D_\tau X_I \gamma_I
-\frac{i}{2}
[X_I,X_J] \gamma_{IJ} \right)\epsilon \n\\
&+&\left(\frac{\mu}{3}\sum_{a=5,6,7} X_a \gamma_a \gamma_{567}
-\frac{\mu}{6}\sum_{s=1,2,3,4,8,9} X_s \gamma_s  \gamma_{567}
\right) \epsilon, \eeqn from which transformation of $\bar \chi$
may also be inferred. Time-dependence of supersymmetry parameter
must be analytically continued as well, \beqn
  \epsilon(\tau) = e^{i {\mu \over 12} \gamma_{567}
  \tau} \epsilon_0.
 \eeqn

\subsection{BPS Bound and Unbroken Supersymmetry}

With a discrete set of vacua, all degenerate with each other, it is
natural to expect an instanton that interpolate between
such vacua. Large number of supersymmetry tends to prohibit
a direct mixing of classical vacua even if an instanton is present,
but it does not mean that instantons do not enter physical processes. 
We will come back to relevancy of instanton in a later section.
These issues will be addressed in section 4 in some detail, but
let us gather some elementary facts about Euclidean equation of
motion.

Gathering bosonic part of Euclidean action that involves the three
bosonic quantity $X^a$'s, 
\beqn 
S_E=\frac 12\int d\tau  \sum_a \Tr
\left\{ (D_\tau X_a)^2 + \left( \frac{\mu}{3}
X^a-i\epsilon^{abc}X^bX^c\right)^2\right\} +\cdots, 
\eeqn 
we may
complete the square one more time, 
\beqn 
S_E&=&\frac 12\int d\tau
\,\sum_a\Tr \left(\mp D_\tau X_a+
 \frac{\mu}{3} X^a-i\epsilon^{abc}X^bX^c\right)^2\nn
&\pm &\int d\tau  \,\sum_a\Tr
\left((D_\tau X_a)\left(\frac{\mu}{3} X^a-
i\epsilon^{abc}X^bX^c\right)\right)\nn
&+&\cdots.
\eeqn
The ellipsis denotes the remainder of the action, which are composed of
either a nonnegative part from the rest of $X^I$'s or fermion bilinears,
none of which would be involved in the interpolating solution.

Setting the other fields to vanish, we find a topological lower
bound of the action from the second line, 
\beqn 
S_E&\ge &\pm \int
d\tau  \,\frac{d}{d\tau}\sum_a\Tr \left(\left(\frac{\mu}{6}
X^aX^a- i\frac{1}{3}\epsilon^{abc}X^aX^bX^c\right)\right)\nn &&=
\pm \Tr\left(\frac{\mu}{6} X^aX^a-
i\frac{1}{3}\epsilon^{abc}X^aX^bX^c)\right)\Biggr\vert^\infty_{-\infty}.
\eeqn 
For configurations interpolating between two distinct vacua, we require
\beqn 
X^a(\infty)&=&-\frac{\mu}{3}J^a_{+} ,\nn
X^a(-\infty)&=&-\frac{\mu}{3}J^a_{-} ,\label{boundary} 
\eeqn 
With this, the lower bound of the Euclidean action collapses to
a trace over the quadratic Casimir invariant, 
\beqn 
S_E&\ge & \pm
\frac{1}{6}\left(\frac{\mu}{3}\right)^3 \left(\Tr
J^a_{+}J^a_{+}-\Tr J^a_{-}J^a_{-}\right) .
\eeqn 
Thus, to find an
instanton solution between a pair of different vacua, it suffices
to solve the first order equations, 
\beq 
\pm D_\tau X^a=
\frac{\mu}{3} X^a-i\epsilon^{abc}X^bX^c \label{BPS} 
\eeq 
under the
boundary condition (\ref{boundary}). The sign is chosen so that
the lower bound for the Euclidean action is nonnegative. It is not
difficult to check that any solution to this equation also solves full
equation of motion.

Any Euclidean solution that solves eq.~(\ref{BPS}) enjoys a
special property that it is preserved by half of 16 dynamical
supersymmetries. Wick rotating the SUSY transformation and
isolating transformation property of the fermions, we find
the following terms  
\beq 
\delta \Psi = \left(
iD_\tau X_a \gamma_a  -\frac{i}{2} [X_a,X_b]\gamma_{ab} +
\frac{\mu}{3}\sum_{a=5,6,7} X_a\gamma_a \gamma_{567} \right)
\epsilon 
\eeq 
on the instanton background. On the other hand, we have
the identity, 
\beq 
\gamma_{ab}=\gamma_c\gamma_{567}\epsilon^{abc},
\eeq 
so that 
\beq 
\delta \Psi = i\left( D_\tau X_a \gamma_a -
\left(\frac{\mu}{3} X_a-i\epsilon^{abc}X^bX^c \right)\gamma_a
(i\gamma_{567}) \right) \epsilon =0 ,
\eeq 
provided that
\beq
i\gamma_{567}\epsilon_0=\pm\epsilon_0 ,
\eeq 
where the sign choice should be correlated with that of (\ref{BPS}).
With each choice of sign, then, these are the 8 preserved
supersymmetries promised. For this reason, we  refer to (\ref{BPS})
as the BPS equation.

\subsection{Instanton Solutions}

An infinite class of instantons can be found with a simplifying
condition, $J^a_+=0$. This gives interpolating configuration
from an arbitrary vacuum at past infinite to the trivial 
perturbative one at
future infinity. The solution is fairly elegant. We use the
ansatz, 
\beq 
X^a=-\frac{\mu}{3}f(\tau)J_-^a ,
\eeq 
with the static
gauge $A_\tau=0$, and look for a solution with $f(-\infty)=1$ and
$f(\infty)=0$. Equations with $-$ sign 
reduces to a single ordinary differential equation, 
\beq
\label{ode} 
\frac{d}{d\tau}f=-\frac{\mu}{3}\left(f-f^2\right),
\eeq
and the unique solution satisfying the boundary condition is, 
\beq
f(\tau)=\frac{1}{1+\exp(\mu (\tau-s)/3)} 
\eeq 
where we kept a
collective coordinate $s$ explicitly to parameterize where the
instanton is centered along the Euclidean time.

The value of the Euclidean action of this interpolating instanton
is 
\beq S_E=\frac{1}{6}\left(\frac{\mu}{3}\right)^3\Tr J^a_-J^a_- .
\eeq 
Writing $J_-$ as sum of irreducible $n_p\times n_p$
representations with $\sum_p n_p =N$, value of the Euclidean
action is 
\beq 
S_E=\frac{1}{6}\left(\frac{\mu}{3}\right)^3\sum_p
\frac{n_p(n_p-1)^2}{4} .
\eeq 
Such an instanton will contribute to some
physical processes but is suppressed by exponentially small factor
\beq e^{-S_E} . \eeq 
Finally, one should remember $g_{YM}^2$ to recover 
the actual {\it dimensionless} exponent is $S_E/g^2_{YM}$.

\section{Zero Modes}

\subsection{Bosonic Zero Modes}

One particular collective coordinate is already manifest in the solution
above, which the Goldstone mode coming from time translation symmetry.
For any finite action solution to the BPS equation $X^a$, the translational
zero mode,
\beq
D_\tau X^a ,
\eeq
is a normalizable zero mode.
A natural followup question is whether instanton solutions have more
bosonic zero modes and whether it is possible to break up an instanton
to two or more intermediate ones.
For this let us perturb the BPS equation. We may assume static gauge again,
$A_\tau=0$, without loss of generality. Denoting by $V^a$ the perturbed
part of $X^a$, we have
\beq
\pm \frac{d}{d\tau} V^a=
\frac{\mu}{3} V^a-i\epsilon^{abc}[X^b,V^c] \label{zero}
\eeq
and the problem reduces to classifying normalizable solutions to this.

The instanton solutions of previous subsection admit relatively
simple solutions to this zero mode counting problem. With the right choice
of sign, the equation is 
\beq 
- \frac{d}{d\tau} V^a= \frac{\mu}{3}
\left(V^a+f(\tau)i\epsilon^{abc}[J^b_-,V^c]\right)  .
\eeq 
In the far
future, $f\rightarrow 0$, and the simple exponential behavior due
to the $\mu/3$ piece is
consistent with the normalizability requirement on zero modes. 
The main issue is then what
happens at past infinity where $f\rightarrow 1$. 

Two technical points are helpful here. First, one may regard
$i\epsilon^{abc}$ as the generators $T^{(b)}_{ac}$ of $SU(2)$. Second,
the commutator action with $J^a$ in a representation $R$ of $SU(2)$
actually form a representation $R\otimes R^*\simeq R\otimes R$. Thus
the matrix operator that must be diagonalized may be thought of as a triple
tensor product
\beq
K\equiv \sum_{b=5,6,7}T^b\otimes (J^b_-\otimes 1 + 1\otimes J_-^b) .
\eeq
Let us simplify further and consider the case of $J_-$ in
the maximal (i.e., spin $(N-1)/2$) irreducible representation.
Square of the $N$ dimensional irreducible representation $[(N-1)/2]$
is decomposed as 
\beq
[(N-1)/2]\otimes[(N-1)/2]=[0]\oplus[1]\oplus\cdots\oplus[N-2]\oplus[N-1] ,
\eeq 
which gives integer spins only. Tensoring each block $[j]$
with $j=0,1,\dots, N-1$ with the adjoint representation [1] of
$T^{(b)}_{ac}= i\epsilon_{abc}$, we find 
\beq 
[1]\otimes
[j]=[j-1]\oplus[j]\oplus[j+1] .
\eeq 
Eigenvalues of the operator $K$
for the three blocks on the right hand side are $-(j+1)$, $-1$,
$j$, respectively. Of these three, only the first can and does
produce normalizable zero mode, $2j-1$ number of them, for $j\ge
1$.

Thus, for an instanton that interpolates between a single maximal
giant graviton in the irreducible $n\times n$ form and the trivial 
vacuum, the bosonic zero modes within the same $n\times n$ block 
may be summarized as 
in the following table,
\vskip 5mm
\begin{center}
\begin{tabular}{c|c|c|c|c|c}
degeneracy & 1 & 3 & $\qquad\cdots\qquad$ &$2N-5$ &$2N-3$ \\ \hline
behavior at $-\infty$ & $e^{(\mu/3)\tau}$ &  $e^{2(\mu/3)\tau}$ &
$\qquad\cdots\qquad$ &  $e^{(N-2)(\mu/3)\tau}$ & $e^{(N-1)(\mu/3)\tau}$
\end{tabular}
\end{center}
\vskip 5mm
All of these zero modes decay as $e^{-(\mu/3)\tau}$ at future infinity.
There are total $(N-1)^2$ number normalizable bosonic zero modes.

Generally speaking,  one might need to take into account of nonnormalizable 
zero modes in case of Euclidean instantons. For solitons, the normalizability 
requirement arises from consideration of low energy effective action
of collective coordinates, since $L^2$ norm of zero modes appear in
the effective kinetic term of the former. For instantons, the only 
requirement is that we maintain the same finite Euclidean action. This
by itself does not restrict the zero mode to be normalizable. 

In particular, there are $N^2-1$ number of zero modes from the above 
(corresponding to $K=-1$) that asymptote to constants at past infinity.
Under such a deformation, the instanton solution remain of the same 
finite action. These zero modes will deform $J_-$ by constant
amount, but since small change that respect the vacuum condition
cannot alter the representation of $SU(2)$, this deformation cannot 
change the action. Combined with the above normalizable zero modes,
we have total $2N(N-1)$ number of possible deformations of a maximal
instanton that preserve the Euclidean action. On the other hand, 
none of the $N^2-1$ nonnormalziable zero modes are physical. They
correspond to rotation of the instanton by global $SU(N)$ unitary 
transformations. That is, these $N^2-1$ modes are generated by
\beq
X^a(\tau)\rightarrow gX^a(\tau) g^\dagger .
\eeq
In the temporal gauge, $A_{10}=0$, we adopted, this is a pure gauge 
transformation when $g$ is uniform. As a metter of definition 
of the theory, such a gauge 
parameter should not be included in the path integral. With more
general instantons with both $J_\pm$ nontrivial, there could be 
relevant nonnormalizable zero modes that rotate one of two $J_\pm$
while leaving the other invariant.

\subsection{Fermionic Zero Modes}

Large number of bosonic zero modes hint at similarly large number of
fermionic zero modes. Here, we will
perform an explicit construction of fermionic zero modes in much the same
way as in bosonic case above.
The fact that there are 8 broken supersymmetry implies existence
of fermionic zero modes arising as Goldstone fermionic modes.
These should be of the form 
\beq 
\psi_0 = i\left( D_\tau X_a \gamma_a
\pm \left(\frac{\mu}{3} X_a-i\epsilon^{abc}X^bX^c \right)\gamma_a
\right) \epsilon'=2i \left(D_\tau X_a \gamma_a\right) \epsilon' ,
\eeq 
with wrong chirality fermions $\epsilon'$; 
\beq
i\gamma_{567}\epsilon_0'=\mp\epsilon_0' 
\eeq 
Below we will also
recover these zero modes from general fluctuation analysis.
Because these arise as Goldstone modes, the associated Grassmann
coordinate are special among fermionic ones. Specifically,
they cannot have an ``interaction vertex'' in the effective action
of the instanton. 

The equation of motion for fermions in the instanton background is
\beq
D_\tau\Psi +\frac{\mu}{4} (i\gamma_{567})\Psi
+ \gamma_a[X^a, \Psi ] 
=0.
\eeq
Let us again specialize to those instantons which interpolate from
giant graviton with representation $J_-^a$ to the trivial vacuum.
Normalizability at $\tau=\infty$ requires $\Psi$ to be of positive
chirality under $i\gamma_{567}$.
Normalizability at $\tau=-\infty$ depends on the last piece, which
on the instanton becomes 
\beq
-\frac{\mu}{3}f(\tau)\gamma^a[J_-^a,\Psi]  ,
\eeq 
so the problem
reduces to an eigenvalue problem under the operator 
\beq 
\tilde K\equiv -\gamma^a\otimes (J_-^a\oplus 1+1\oplus J_-^a) .
\eeq 
A zero
mode is normalizable if and only if the eigenvalue of this
operator is negative and $<-3/4$.

Because only 3 out of 9 possible $\gamma$'s are involved, this
can also be solved using $SU(2)$ algebra. Let us write the three 
gamma matrices
in a complex form
\beq
\gamma^a=\sigma^a\otimes\Gamma ,
\eeq
where $\Gamma$ is some chirality operator such that $\Gamma^2=1$.
We will use the fact that $i\gamma_{567}=-1\otimes\Gamma$, 
meaning that the eigenvalue of $\Gamma$ is negative of that of
$i\gamma_{567}$. Since normalizable zero modes must have positive
chirality under the latter, we have in effect,
\beq
\tilde K\equiv \sigma^a\otimes (J_-^a\oplus 1+1\oplus J_-^a) ,
\eeq
where we should remember the extra degeneracy coming from extra
spinor indices acted on by $\Gamma$. After taking into account that 
$\Gamma=-1$, this degeneracy gives a factor of 4 when counting real
fermionic collective coordinates.

As in bosonic case, we consider the maximal irreducible instanton of size
$N\times N$ for $J_-$ side initially. Square of the $N$-diemsnional 
representations is 
\beq
[(N-1)/2]\otimes[(N-1)/2]=[0]\oplus[1]\oplus\cdots\oplus[N-2]\oplus[N-1]
\eeq 
Tensoring each block $[j]$ with $j=0,1,\dots, n-1$ with a
spin half representation [1/2] of $\sigma/2$, we find 
\beq
[1/2]\otimes [j]=[j-1/2]\oplus [j+1/2] 
\eeq 
eigenvalues of the
operator $\tilde K$ for the three blocks on the right hand side are
$-(j+1)$ and $j$, respectively. Of these two, only the first can
and does produce normalizable zero modes, $2j$ number of them, for
$j\ge 1$.
Thus, for an instanton that interpolates between a single giant graviton
in an irreducible $n\times n$ block and the trivial vacuum, the fermionic
zero modes may be summarized as in the following table,
\vskip 5mm
\begin{center}
\begin{tabular}{c|c|c|c|c|c}
degeneracy & $4\times 2$ & $4\times 4$ & $\qquad\cdots\qquad$
&$ 4\times(2N-4)$
&$4\times(2N-2)$ \\ \hline
behavior at $-\infty$ & $e^{(5\mu/12)\tau}$ &  $e^{(3\mu/4)\tau}$ &
$\qquad\cdots\qquad$ &  $e^{(4N-7)(\mu/12)\tau}$ & $e^{(4N-3)(\mu/12)\tau}$
\end{tabular}
\end{center}
\vskip 5mm
All of these zero modes decay as $e^{-(\mu/4)\tau}$ at future infinity.
Thus, given an irreducible $N$-dimensional $J_-$, The total number of 
real fermionic collective coordinates from these zero modes within the
same $N\times N$ block is $4N(N-1)$.\footnote{
An unusual fact is that this number of fermionic zero modes match
up with the bosonic side, only if we inlcude all $N^2-1$ nonnormalizable
zero modes in the latter. The latter gives $(N-1)^2+(N^2-1)=
2N(N-1)$, so one real bosonic mode correponds to one complex fermionic
mode. The reason behind this must be that 
supersymmetry parameter itself has a exponential dependence on 
$\tau$, although it is not clear to us whether there is a precise
one-to-one matching between the two sets.}

Of these the first 8 modes are associated with 8 broken supersymmetry.
The Goldstone modes of the latter  are proportional to
\beq
iD_\tau X^a\gamma^a e^{(\mu/12)\tau}\epsilon_0'
\eeq
with $(i\gamma_{567})\epsilon_0' =\epsilon_0'$. Comparing the asymptotic
form of this to the first 8 zero modes, we can see easily that they are
of identical form.

\subsection{A Brief Comment on Stability of Vacua}

These 8 fermionic zero modes associated with the broken
supersymmetry are quite prohibitive in that instantons 
amplitudes get further suppression due to them.
Generally speaking instanton will
contribute to an operator ${\cal O}$ via a path integral, 
\beq
\int [dX][d\Psi]\,{\cal O}\,e^{-S} .
\eeq 
Essential part of such a
computation comes from integration over bosonic and fermionic
collective coordinates, call them $z^i$ and $\zeta^l$, whereby
the path integral includes a piece 
\beq 
\int \prod_i dz^i\prod_l
d\zeta^l \,{\cal O}\,e^{-S_E-\delta S(z^i,\zeta^l)}  
\eeq 
Integral
over Grassmann numbers $\zeta^l$ vanishes unless the integrand
produces matching fermions in one-to-on fashion.

For most fermionic zero modes, this is easy to arrange regardless
of ${\cal O}$, since the higher order correction to Euclidean
action, $\delta S(z^i,\zeta^l)$, will generally have a potential
term involving four (or more) $\zeta^l$'s. This does not contradict
the statement that $\zeta^l$ are zero modes, since the latter
simply implies absence of quadratic terms in case of fermions.
Such quartic potential term for fermionic collective coordinates
has been derived and used extensively in the context of Yang-Mills
theory.\footnote{Direct derivation of such higher order terms with
fermionic collective coordinate is 
particularly well-documented for the case of low energy dynamics
of monopoles in 4 dimensions. See Ref.~\cite{gauntlett} for detailed
description of how such terms arise.}

Exception to this are Goldstone zero modes, bosonic or fermionic,
representing spontaneously broken global symmetries. Nonvanishing 
amplitude is possible only if one starts with ${\cal O}$ with enough 
fermions in it to cancel Goldstone fermionic coordinates. It
shows that there is no
mixing among classical vacua by these instantons, and that each
classical vacuum of giant gravitons must remain robust against quantum
tunneling. Effect of an instanton will show up only at higher order
operators. 

This, together with stability 
argument of Ref.~\cite{jhp}, implies
that the classical vacua of giant
gravitons are in fact exact quantum mechanical vacua.
One could argue that being protected states 
is by itself enough to guarantee that classical supersymmetric
vacua carry over to quantum mechanics, yet, this sort of 
argument is known to fail in general supersymmetric theories.
Such stability argument applies within full quantum theory which
is in principle distinct from classical theory no matter how
close one gets near the latter. 
There are known cases where, despite classical 
BPS state, no quantum counterpart exist no matter how close to
classical limit one takes \cite{su3}. The present theory seems to
escape such a possibility, likely due to relatively
large number of supersymmetries.

\section{General Moduli Spaces}

So far, we studied a rather restricted class of
instantons (with $J_+=0$), and counted zero modes of even more 
special case of them, namely those with $J_-$ in the irreducible
$N$-dimensional representation filling out the entire $U(N)$.
We found $(N-1)^2$ normalizable bosonic zero modes, $N^2-1$
nonnormalizable one which are unphysical due to gauge symmetry,
and $4N(N-1)$ (real) fermion zero modes. As with any quantum theory,
one must understand entire species and moduli spaces of instantons
to make proper usage of it in path integral, so it is of prime
importance to catalog general instantons and their moduli spaces.

For instance, we should ask how the counting changes if we
instead consider $n\times n$ irreducible instanton embedded in
$U(N)$ theories. Bosonic zero mode arising from the same 
$n\times n$ block should follow from the above zero mode counting,
so we have $(n-1)^2$ physical modes and $n^2-1$ unphysical gauge
modes. Repeating the anaylsis of the previous section for ``off-diagonal''
part of matrices $X^a$'s, we find additional modes. Of these,
$2(N-n)(n-2)$ are normalizable and have to be physical. In addition, 
$2(N-n)n$ modes that approach constant at $\tau=-\infty$ also exist
but these are all accounted by global gauge rotation that takes
the $n\times n$ solution out into entire $N\times N$. We could
repeat this exercise for more general case with more than one
irreducible blocks embedded inside $U(N)$;
it is again matter of studying eigenmodes of the algebraic 
operator $K$, which become effectively
\beq
K_{ij}\equiv \sum_{b=5,6,7}T^b\otimes (J^b_i\otimes 1 + 1\otimes J^b_j) ,
\eeq
on different blocks in $U(N)$ when $J_{i}$'s are independent 
irreducible representations of $SU(2)$ inside $J_-$.

For classification of instantons and their moduli space, it would
be desirable to approach the problem with more global viewpoint.
It turns out that  this issue was studied
in depth previously. An essentially same system of equations was
considered before in a different setting of ${\cal N}=1^*$ $U(N)$
Yang-Miils theory \cite{Bachas}. It just so happens that the
vacuum condition and BPS equation for domain walls in ${\cal N}=1*$
theories are identical
to our vacuum condition and BPS equation for instantons. They
further mapped this system of equations into a mathematical
problem of an infinite dimensional hyperK\"ahler quotient procedure
and constructed the solution space of the domain walls.

Among the information one could extract from their general approach 
would  be which domain wall (instanton) may split into which combination of 
domain walls (instantons). On the other hand, counting of deformation
was such that some pure gauge mode are counted along with physical 
ones.\footnote{Notably, this is the case when $J_+=0$ \cite{Bachas}. }
Indeed numbers in Ref.~\cite{Bachas} match precisely with ours whenever the
comparison is possible, once we include the global gauge modes in the
counting. The constructive approach we took is more detailed in that it
gives a simple distinction between the two classes of deformations.
Thus, when carefully combined, the two approaches should generate
essentially all information of the physical moduli space of instantons,
which would be vital if we were computing instanton effect via the
Eulidean path integral. In this note, we will not attempt this
general anaysis.

\section{Vacua and Instantons in the Presence of Fivebranes}

A mass-deformation of ${\cal N}=8$ Matrix model including
hypermultiplets was derived in \cite{kimleeyi}. It was constructed to
describe longitudinal fivebranes in the pp-wave background.
It was also found that there exist many continuous family of
classical vacuum solutions which preserve full dynamical
symmetries. Here, we will investigate whether
there are instanton solutions interpolating between these vacua.

To set up the notation, let us briefly recall what new fields are
involved in this ${\cal N}=8$ quantum mechanics. When we have $k$
longitudinal fivebranes probed by $U(N)$ M(atrix) theory, the only
modification to the $U(N)$ quantum mechanics is introduction of
$k$ hypermultiplets in the fundamental representations. This
necessarily breaks SUSY down to ${\cal N}=8$ from ${\cal N}=16$,
and thus split the adjoint Yang-Mills fields also into two ${\cal
N}=8$ multiplets. In the BMN deformed M(atrix) theory considered
above, 9 adjoint scalars are already split into 3+6. We labeled
the first 3 as $X^{5,6,7}$. In this notation, $X^{6,7,8,9}$ fall
under the adjoint hypermultiplet. We further combine these 4
scalars in a complex form, 
\beq 
y_2=X^6-iX^7,\qquad
y_1=X^8-iX^9.
 \eeq 
 Whenever we introduce an longitudinal fivebrane
sitting at the center, we should add a fundamental hypermultiplet 
with two complex scalars, 
\beq 
q_2^{(f)},\qquad q_1^{(f)}.
\eeq 
Our convention is such that $q$'s are
anti-fundamental representations while their conjugates, $\bar q$
and $\bar q$, are in the fundamental representation.

We should note here that Ref.\cite{kimleeyi} found two distinct
class of vacuum solutions invariant under all supersymmetries. One
set involving $X^5,y_2, q_2$ give obvious generalization of the
Giant graviton in the bulk, while the other involves $X^5, y_2,
q_1$. In the following we found instanton solutions associated
with the former only. There appears to be no supersymmetric instanton
solution associated with the latter.

\subsection{Abelian Case}

For simplicity, we first consider the instanton solution interpolating
between an Abelian giant graviton at past infinity and trivial
vacuum at future infinity.
Turning off irrelevant fields, we write down the Euclidean action
in a complete square form,
\beqn
 S_E &=& {1 \over 2} \int d \tau  {\rm Tr} \left\{ \left(\mp D_\tau X_5 +
    \left({1\over 2} \sum_f \bar q_2^{(f)} q_2^{(f)}
        + {\mu \over 3} X_5 \right) \right)^2  \right. \n\\
    &{}& \qquad  \left. + \sum_f
       \left(\mp D_\tau \bar q_2^{(f)} +
        \left(X_5 + {\mu \over 3}\right) \bar q_2^{(f)} \right)
       \left(\mp D_\tau  q_2^{(f)} +  q_2^{(f)} \left(X_5
        + {\mu \over 3}\right)  \right) \right\} \n \\
     &{}& \pm {1 \over 2} \int d \tau {\rm Tr}
      \left\{ \left(2 D_\tau X_5 \right)
      \left({1\over 2} \sum_f \bar q_2^{(f)} q_2^{(f)}
      +  + {\mu \over 3} X_5 \right)  \right. \n\\
     &{}& \qquad \left. +\sum_{(f)} D_\tau \bar q_2^{(f)}
      \left(X_5 + {\mu \over 3} \right)
     q_2^{(f)} + \sum_f \left(X_5 + {\mu \over 3} \right)  \bar q_2^{(f)}
     D_\tau q_2^{(f)} \right\} \n \\
  &\geq& \pm {1\over 2} \int d \tau {d \over d\tau} {\rm Tr}
     \left({\mu \over 3} X_5^2 + {\mu \over 3}\sum_f \bar q_2^{(f)} q_2^{(f)}
       + X_5\sum_f \bar q_2^{(f)} q_2^{(f)} \right) \n \\
   &=& \pm{1 \over 2} {\rm Tr} \left( {\mu \over 3} X_5^2
      + {\mu \over 3}\sum_f \bar q_2^{(f)} q_2^{(f)}
       + X_5\sum_f \bar q_2^{(f)} q_2^{(f)}
    \right) \Biggr\vert_{-\infty}^{+\infty}.
\eeqn
Lower bound is saturated if the first order equations are satisfied,
\beqn
 \pm D_\tau X_5 &=& {1\over 2} \sum_f \bar q_2^{(f)} q_2^{(f)}
             + {\mu \over 3} X_5 ,\n \\
 \pm D_\tau \bar q_2^{(f)} &=& X_5  \bar q_2^{(f)}
  + {\mu \over 3} \bar q_2^{(f)}.
\eeqn
Dropping time dependence, we recover a Giant graviton of the form
\beqn
 X_5 = -{\mu \over 3}, \qquad
 q_2^{(f)}= \sqrt{2} {\mu \over 3} z^f,
\eeqn
where $z$ is any complex $k$-vector of unit length. Vacuum moduli space
is $CP^{k-1}$.

The ansatz for the instanton solution
interpolating between an Abelian giant graviton at past infinity
and perturbative vacuum at future infinity is
\beqn
 X_5 (\tau) = - {\mu \over 3} f(\tau), \qquad
  \bar q_2^{(f)} (\tau) = \sqrt{2} {\mu \over 3} z^f g(\tau),
\eeqn
with $f(-\infty)=1=g(-\infty)$ and $f(+\infty)=0=g(+\infty)$.
Inserting it into the first order equations with $-$ sign,
we get
\beqn
 \dot{f} = - {\mu \over 3} \left(f - g^2 \right) ,\qquad
 \dot{g} = - {\mu \over 3} \left(g - fg \right).
\eeqn
A unique solution to this with the right asymptotic behavior is 
(\ref{ode})
\beqn
 f=g={1 \over {1+ \exp(\mu (\tau-s)/3)}} ,
\eeqn
whose form is identical to the above ${\cal N}=16$ case.
The action is 
\beqn
 S_E= {1 \over 2}
 \left({\mu \over 3}\right)^3.
\eeqn
This solution has a single bosonic zero mode which corresponds
to a Goldstone mode.

\subsection{General Case}

We now repeat the exercise for non-Abelian case and consider instanton
solutions interpolating between a nonabelian giant graviton at past
infinity and trivial vacuum at future infinity. Turning on adjoint
scalars $X_5$, $\bar y_2$ and fundamental scalars $\bar q_2^{(f)}$
only, we have the following inequality of Euclidean action,
\beq
 S_E
  \geq \pm{1 \over 2} {\rm Tr} \left( {\mu \over 3} X_5^2
      + {\mu \over 3}\sum_f \bar q_2^{(f)} q_2^{(f)}
       + {\mu \over 3} \bar y_2 y_2
       + X_5\sum_f \bar q_2^{(f)} q_2^{(f)}  + X_5 \left[\bar y_2, y_2 \right]
    \right) \Biggr\vert_{-\infty}^{+\infty}.
\eeq
The bound is saturated if and only if
\beqn
\label{x5}
 \pm D_\tau X_5 &=& {1\over 2} \sum_f \bar q_2^{(f)} q_2^{(f)}
             + {\mu \over 3} X_5 + {1 \over 2} \left[\bar y_2, y_2 \right]  ,\\
\label{q2}
 \pm D_\tau \bar q_2^{(f)} &=& X_5  \bar q_2^{(f)}
  + {\mu \over 3} \bar q_2^{(f)}  , \\
  \label{y2}
 \pm D_\tau \bar y_2 &=& \left[ X_5 , \bar y_2 \right] + {\mu \over 3}
  \bar y_2.
\eeqn
For the instanton solution interpolating between
an nonabelian giant graviton occupying an $n\times n$ block of
adjoint scalars at past infinity and perturbative vacuum
at future infinity,  we use an ansatz of the form,
\beqn
 - (X_5)_{AB}(\tau) &=& {\mu \over 3} (m+1-A) f(\tau) \delta_{AB}, \n \\
 (\bar q_2^{(f)})_A(\tau) &=&
   \sqrt{2} {\mu \over 3} \omega_2^f  g(\tau)\delta_{A,m},  \n \\
  (\bar y_2)_{AB}(\tau) &=&
   \sqrt{2} {\mu \over 3} \alpha_A h(\tau) \delta_{A,B-1},
\eeqn 
where
\beqn 
|\alpha_{A}|^2 -|\alpha_{A-1}|^2 = m+1-A  ,
\eeqn 
for all $1\le A\le  n$ except for $A=m$ \beq
\sum|w^f_2|^2+|\alpha_m|^2 -|\alpha_{m-1}|^2 = 1 , \eeq Here  $m$
is any fixed integer such that $(n-1)/2 < m\le n$. 

To find an instanton that interpolate between a giant graviton and
perturbative vacua, we set the boundary conditions,
\beqn
&&f(-\infty)=g(-\infty)=h(-\infty)=1 ,\n \\
&&f(+\infty)=g(+\infty)=
h(+\infty)=0 .
\eeqn
Taking an appropriate sign choice, and writing 
equation for $X^5$ component-wise, we find
\beq
\dot{f} = -{\mu \over 3} \left( f - h^2 \right) ,
\eeq
from all diagonal entries except for $A=m$ case. The latter gives
a slightly different equation
\beq
\dot{f} = -{\mu \over 3}
     \left( f - h^2\right) +\frac{\mu}{3}|\omega_2|^2 (g^2 -h^2) .
\eeq
Two equations are consistent only if $g^2=h^2$, 
which in turn forces $g=h$.
Then equations for $y_2$ and for $q_2$ collapse to one,
\beq
\dot h= -\frac{\mu}{3}\left(h-fh\right) .
\eeq
Unique solution consistent with the boundary condition is
\beqn
 f=g=h={1 \over {1+ \exp(\mu (\tau-s)/3)}}.
\eeqn
Again, dependence on $\tau$ is identical to ${\cal N}=16$ 
case. Value of the Euclidean action also has a simple expression, 
\beq 
S_E\ge - {1 \over 2} \left({\mu \over 3} \right)
{\rm Tr} (X_5)^2 \Biggr\vert_{-\infty}^{+\infty} . 
\eeq 
This is
essentially of the same form as the Euclidean action in ${\cal
N}=16$ theory. With large $N$, it again scales as $\sim (\mu
n)^3$.

\subsection{Supersymmetry}

As with ${\cal N}=16$ case, instantons of ${\cal N}=8$
theories also preserves some supersymmetry. Euclideanization of
the theory and the supersymmetry
proceed similarly, and we have
\beqn
\delta \lambda_1 &=& i D_\tau X_\mu \gamma_\mu
\epsilon_1 -i \sum_{\mu , \nu}^5 \frac{1}{2} [X_\mu,X_\nu]
\gamma_{\mu  \nu}\epsilon_1  +iD^3\epsilon_1 + (iD^1+D^2)\epsilon_2 \n\\
&&-i\frac{\mu}{6} \sum_{\mu=1}^4
X_\mu\gamma_\mu \gamma_5 \epsilon_1 +i \frac{\mu}{3} X_5 \epsilon_1,\n \\
\delta \chi_1 &=&-iD_\tau y_i \epsilon_i - i \sum_{\mu=1}^5[X_\mu,
y_i]\gamma_\mu \epsilon_i + \frac{i\mu}{6}\gamma_5 y_1 \epsilon_1 +
\frac{i\mu}{3} \gamma_5 y_2 \epsilon_2  ,\n\\
\delta \psi_1 &=& -iD_\tau  q_i \epsilon_i + i \sum_{\mu=1}^5
q_iX_\mu \gamma_\mu \epsilon_i + \frac{i\mu}{6}\gamma_5 q_1
\epsilon_1 + \frac{i\mu}{3} \gamma_5 q_2 \epsilon_2  ,
\eeqn 
and
similar expressions for $\delta\lambda_2$, $\delta\chi_2$, and
$\delta\psi_2$. $D^{1,2,3}$ are familiar auxiliary fields of
vector multiplet. $\lambda_i$, $\chi_i$, and $\psi_i$  are
4-component Dirac spinors from vector multiplet, adjoint
hypermultiplet, and fundamental hypermultiplets.\footnote{ In
Minkowski signature, they are each symplectic Majorana such that
\beq 
\lambda_1= -i\tilde B \lambda_2^*, \qquad \chi_1=+i\tilde B
\chi_2^*, \qquad \psi_1=+i\tilde B \psi_2^* 
\eeq 
with some
conjugation matrix $\tilde B$. Similarly, supersymmetry parameter
$\epsilon_i$ are symplectic Majorana spinors, 
\beq 
\epsilon_1=
-i\tilde B \epsilon_2^*, \qquad 
\eeq 
Once Euclideanized, however,
this Majorana condition cannot be imposed and we must treat two
components under $i$ index independently. As before, correct
degrees of freedom is found by giving up hermiticity and treating
the path integral as if it is a contour integral.} To recover
$\delta\lambda_2$, $\delta\chi_2$, and $\delta\psi_2$, one must
care to Euclideanize {\it after } performing charge conjugation.
For the instantons in question, this reduces to 
\beqn 
\delta
\lambda_1 &=& +i D_\tau X_5 \gamma_5
\epsilon_1 +iD^3\epsilon_1 +i \frac{\mu}{3} X_5 \epsilon_1,\n \\
\delta \chi_1 &=&-iD_\tau y_2 \epsilon_2 - i [X_5,
y_2]\gamma_5 \epsilon_2  +
\frac{i\mu}{3} \gamma_5 y_2 \epsilon_2 , \n\\
\delta \psi_1 &=& -iD_\tau  q_2 \epsilon_2 + i q_2X_5 \gamma_5
\epsilon_2 + \frac{i\mu}{3} \gamma_5 q_2 \epsilon_2  .
\eeqn 
On
shell, the value of $D^3$ is 
\beq 
D^3= {1\over 2} \sum_f \bar
q_2^{(f)} q_2^{(f)}+ {1 \over 2} \left[\bar y_2, y_2 \right]  .
\eeq
Instanton solutions then preserves half of supersymmetry.
Preserved supersymmetry are generated by $\epsilon^+_1$ and
$\epsilon_2^-$ where 
\beq 
\gamma_5\epsilon_i^\pm =\pm 
\epsilon_i^\pm  ,
\eeq 
which also imply existence of 4 Goldstone
fermionic zero modes generated by the other half $\epsilon_1^-$
and $\epsilon_2^+$. This gives rise to 4 real Grassmannian
collective coordinates, which is free of any interaction vertex.
While larger instanton solutions will admit more zero modes,
bosonic and fermionic, this 4 plus a single bosonic coordinate
from translational symmetry will remain special in that they
remain free no matter what.

\section{Instantons in $AdS \times S$ and Penrose Limit}

pp-wave background in eleven dimension is obtained from $AdS_7
\times S^4$ or $AdS_4 \times S^7$ by taking the Penrose limit.
Instantons interpolating between giant graviton solutions in $AdS
\times S$ spaces were found in \cite{gginads}. We take Penrose
limit of $AdS \times S$ space and compare the limiting form of
instantons to those in the BMN M(atrix) theory.

 First, We briefly summarize Penrose
limit of $AdS_7 \times S^4$ and $AdS_4 \times S^7$. Global metric
for $AdS_7 \times S^4$ is
\beqn 
&&F_4 = 3 R_S^3 {\rm vol} (S^4) \n \\
&& ds^2 = R_{AdS}^2 \left( - \cosh^2 \rho d\tau^2 + d\rho^2 + \sinh^2 \rho
            d \Omega_5^2 \right) + R_S^2 dS_4^2,
\eeqn
where
\beqn
 dS_n^2 = d\theta_n^2 + \sin^2\theta_n^2 dS_{n-1}^2.
\eeqn 
Here $R_{AdS} = 2 R_S = 2 l_p (\pi N)^{1 \over 3}$ and $l_p$ is
eleven dimensional Planck length.  Penrose limit is taken 
with $R_S \rightarrow \infty$ ( or $N \rightarrow \infty$ ) following  
the reparametrizations 
\beqn
 && \theta_i = {\pi \over 2} - {y_i \over R_S} \quad (i=1,2,3) , \n \\
 && \rho = {z \over R_{AdS}},  \n\\
 && \tau = {\mu x^+ \over 6} + {6 x^- \over R_{AdS}^2 \mu},  \n\\
 && \theta_4 = {\mu x^+ \over 3} - {3 x^- \over R_S^2 \mu}.  
\eeqn 
Finally we get the plane wave geometry 
\beqn 
\label{penrose}
 && ds^2 = -4 dx^+dx^- - \left( \left({\mu \over 3}\right)^2 y^2
  + \left({\mu \over 6}\right)^2
       z^2 \right) dx^{+2} +
  d \vec{y}^2 + d\vec{z}^2,  \n\\
 && F_4 = \mu dx^+ \wedge dy^1 \wedge dy^2 \wedge dy^3.
\eeqn

Global metric for $AdS_4 \times S^7$ is
\beqn
 && ds^2 = R_{AdS}^2 \left( - \cosh^2 \rho d\tau^2 + d\rho^2 + \sinh^2 \rho
            d \Omega_2^2 \right) + R_S^2 dS_7^2,  \n \\
 && F_4 = 3 vol(AdS_4).  
\eeqn
Here $R_{AdS} = 1/2  R_S = 1/2 l_p (32 \pi^2 N)^{1\over 6}$.
Again Penrose limit is taken with the limit $R_S \rightarrow \infty$ 
after the reparametrizations 
\beqn
 && \theta_i = {\pi \over 2} - {z_i \over R_S} \quad (i=1,\cdots, 6),  \n \\
 && \rho = {y \over R_{AdS}},  \n\\
 && \tau = {\mu x^+ \over 3} + {3 x^- \over R_{AdS}^2 \mu},  \n\\
 && \theta_7 = {\mu x^+ \over 6} - {6 x^- \over R_S^2 \mu}. 
\eeqn
We get the same geometry as
(\ref{penrose}).

 Instanton solutions interpolating between perturbative vacuum and
spherical membranes in pp-wave background have two-fold origins in
$AdS \times S$ space: (1) One coming from instantons interpolating
between perturbative vacuum and rotating spherical brane in $AdS_4$ 
of $AdS_4 \times S^7$ and (2) the other coming from instantons
interpolating between perturbative vacuum and  rotating spherical brane in
$S^4$ of $AdS_7 \times S^4$. The form of solutions are presented
in \cite{gginads}. Taking Penrose limit of case (2), we find that the 
limiting form of instanton is identical
to that obtained in Matrix theory : 
 \beq
  y = {R_s p \over 1 + \exp(\mu\tau /3)}, 
 \eeq
where $p$ is Penrose limit of angular momentum defined in
\cite{gginads}. Instanton interpolates between $y_0=0$ and $y_p=
R_s p$. Value of the Euclidean action is
 \beqn
  S_E \approx {1 \over 6} N p^3 = {R_s^3 p^3 \over 6 \pi l_p^3}
  = {y_p^3 \over 6 \pi l_p^3} .
 \eeqn
For the case of (1), the Penrose limit of the instanton asymptotes to
 \beqn
  y \sim  { R_{AdS} \,\, \tilde p \over {1 + \exp(\mu  \tau/3) }}.
 \eeqn
$\tilde p$ is again angular momentum of giant graviton in the $AdS$ space. 
Value of the Euclidean action,
 \beqn
  S_E \approx {\sqrt{2} \over 12 } \sqrt{N} \tilde p^3
  = {R_s^3 \over 48 \pi l_p^3} \tilde p^3 =
  {y_{\tilde p}^3 \over 6 \pi l_p^3},
 \eeqn
so we have the same value of the action both cases. This is
consistent with the fact that we have the same pp-wave limits for
both $AdS_7 \times S^4$ and $AdS_4 \times S^7$.

As an aside, we may 
consider instanton solutions which interpolate between
perturbative vacuum and spherical fivebranes. Again we note that 
$S^5$ giant gravitons could arise from two sources: 
(3) fivebranes in $S^7$ of $AdS_4 \times S^7$ and 
(4) fivebranes in $AdS_7$ of $AdS_7 \times S^4$.
For the case (3), the instanton after taking Penrose limit has the
form
 \beqn
  z^4 = {R_s^4 p \over 1+ \exp(\mu \tau/6) }.
 \eeqn
while the Euclidean action is evaluated to be
 \beqn
  S_E \approx {1\over 3} N p^{3 \over 2} ={R_s^6 \over 96 \pi^2
  l_p^6} p^{3 \over 2}
   ={z_p^6 \over 96 \pi^2 l_p^6}.
 \eeqn
For the case (4), the solution again takes the same asymptotic
form as in case (3). Value of Euclidean action is again,
 \beqn
  S_E \approx {2 \over 3} N^2 \tilde p ^{3 \over 2} = {2 R_s^6
  \over 3 \pi^2 l_p^6} \tilde p^{2 \over 3} = {z_{\tilde p}^6
  \over 96 \pi^2 l_p^6}.
 \eeqn
As with $S^2$ case, 
we get the same value of the actions for instantons of different
origins. Note that $S_E \sim x^3$ for instantons interpolating
between membranes and $S_E \sim x^6$ for instantons interpolating
between fivebranes. The values of action just count the
corresponding Euclidean volume of bubble creation and agree 
with the dimensional analysis. Recently it was proposed \cite{malram} 
that $S^5$ giant 
graviton vacua and $S^2$ giant graviton vacua are dual to each other 
in this Penrose limit. This would imply that the two classes
of instantons we considered above should be related to
each other in the Penrose limit. However, the hypothesis 
is not amenable to the naive interpolation of $S^5$ 
giant graviton from $AdS\times S$ to the Penrose limit.
This fact is also clear from behavior of the instanton
solution above. If there is such a correspondence, it is 
apparently not visible in a semiclassical approach, as was 
noted in Ref.~\cite{malram}.

\section{Summary}

In this note, we studied Euclideanized 
 BMN M(atrix) theory
and constructed instanton solutions that interpolate between 
classically distinct vacua. We considered two kinds of M(atrix) theory.
One is the original  BMN M(atrix) theory with ${\cal N}=16$ supersymmetries, 
while the other with reduced ${\cal N}=8$ supersymmetry corresponds
to having longitudinal fivebranes in the background. 

We gave equations for supersymmetric instantons, and solved them for all
possible instantons connecting to the trivial perturbative vacuum. We gave 
detailed counting and, in some cases, explicit construction of zero 
modes around such solutions. This includes Goldstone modes.
and due to some of fermionic Goldstone modes, instantons
cannot mediate mixing of vacua and each classical vacuum remain
robust against quantum tunneling. 
Instantons will certainly contribute to higher order interaction 
vertices, nevertheless. It remains an open question as to 
what specific
roles instantons will play in uncovering quantum nature of BMN 
M(atrix) theory.

\subsection*{Acknowledgement}

We are  grateful to Jeong-Hyuck Park and O-Kab Kwon for useful discussions. 
We also thank B. Pioline for drawing our attention to Ref.~\cite{Bachas}.
PY is supported in part by Korea Research Foundation (KRF)
Grant KRF-2002-070-C00022. JTY is supported in part by Stichting FOM.

\end{document}